\documentclass{article}

\usepackage{arxiv}

\usepackage{cite}
\usepackage{amsmath,amssymb,amsfonts}
\usepackage{algorithmic}
\usepackage{graphicx}
\usepackage{textcomp}
\usepackage{xcolor}
\usepackage{url}
\usepackage{booktabs}
\usepackage{multirow}
\usepackage{listings}
\title{Automated BPMN Model Generation from Textual Process Descriptions: A Multi-Stage LLM-Driven Approach}

\author{
    {\hspace{1mm}Ion Matei} \\
	Fujitsu Research of America\\
	\texttt{imatei@fujitsu.com} \\
	\And  
    {\hspace{1mm}Maksym Zhenirovskyy} \\
	Fujitsu Research of America\\
	\texttt{mzhenirovskyy@fujitsu.com} \\
    \And
    {\hspace{1mm}Praveen Kumar Menaka Sekar} \\
	University of Maryland\\
	\texttt{praveenm@umd.edu} \\   
    \And
    {\hspace{1mm}Hon Yung Wong} \\
	Fujitsu Research of America\\
	\texttt{awong@fujitsu.com} \\      
}

\date{}

\renewcommand{\headeright}{}
\renewcommand{\undertitle}{}
\begin{document}
\maketitle

\begin{abstract}
Automatically reconstructing BPMN models from unstructured natural-language descriptions remains challenging due to heterogeneous modeling conventions, multilingual sources, and the lack of reliable ground truth. We present a scalable, multi-stage LLM-driven pipeline that automates both ground-truth construction and model reconstruction. Multilingual BPMN XML files are translated into English, validated using execution-oriented compliance checks in SpiffWorkflow, and iteratively repaired through targeted LLM-guided corrections to produce a consistent ground-truth corpus. From these validated models, process descriptions are generated and used to reconstruct executable BPMN~2.0 XML diagrams without manual curation. We introduce a multi-dimensional similarity framework combining structural metrics, type-distribution alignment, and embedding-based semantic measures. In an empirical study of 750 public BPMN diagrams, the pipeline generated 387 validated ground-truth models and achieved average reconstruction similarity above 0.75, including approximately 50 near-perfect reconstructions differing only in minor naming variations. The results demonstrate that LLMs can generate structurally compliant and semantically meaningful BPMN diagrams at scale.
\end{abstract}

% keywords can be removed
\keywords{process Modeling \and BPMN \and LLMs \and process mining \and workflow automation}

\section{Introduction}
\label{sec:introduction}

Modeling policies in Business Process Model and Notation (BPMN) \cite{omg2011bpmn} enables analysis, execution, monitoring, and optimization. For example, BPMN diagrams encode clinical pathways understandable to clinicians while supporting tool integration \cite{kassim2022bpmn}, and enable resource optimization in hospital settings \cite{10.1007/978-3-030-04771-9_12}. However, much process knowledge exists only in textual form, and existing extraction methods often produce models that are syntactically invalid or non-executable. We address this limitation with a systematic, multi-stage LLM pipeline that incrementally builds process understanding and generates fully compliant  BPMN diagrams.

Our first contribution is an automated ground-truth construction pipeline that standardizes heterogeneous models and enforces, in part,  execution compliance. We translate publicly available BPMN files from multiple languages to English while preserving XML structure, apply formal validation using SpiffWorkflow \cite{spiffworkflow} to detect compliance issues, and integrate an LLM-guided correction loop to repair non-compliant models at scale. This produces a corpus of corrected models suitable as ground truth, from which we automatically generate process descriptions using LLMs.

Our second contribution is an LLM-based reconstruction method that mimics human modelers. The pipeline extracts salient process elements and decision logic, constructs data models and activity–data mappings, synthesizes BPMN~2.0 XML under explicit model checker requirements (e.g., default paths for exclusive gateways, condition expressions for non-default branches).

Our third contribution is a multi-dimensional similarity framework integrating: (i) topological correspondence via graph statistics and degree-sequence correlation, (ii) type-distribution similarity using Jensen–Shannon divergence, (iii) semantic similarity from sentence-level embeddings with optimal assignment, and (iv) contextual structural–semantic variants including local neighborhood features. We evaluate via an empirical study starting with 750 public BPMN diagrams, yielding roughly 400 validated ground-truth models after translation and correction. Despite the absence of human curation, the method achieves an average similarity exceeding 0.75, with approximately 50 near-perfect reconstructions differing only in minor naming variations.

\textit{Paper structure}: Section~\ref{sec:ground truth} details the ground-truth dataset generation methodology. Section~\ref{sec:bpmn generation} introduces the multi-stage BPMN generation framework. Section~\ref{sec:comparison} outlines the evaluation framework. Section~\ref{sec:experimental results} presents our empirical study, while Section~\ref{sec:related work} sets our contributions within context of existing research.

\section{Ground truth generation}
\label{sec:ground truth}

To evaluate the reconstruction pipeline, we require BPMN models paired with process descriptions. We construct this dataset automatically from publicly available BPMN files, which cannot be assumed compliant and are infeasible to verify manually. The methodology consists of three stages: (i) multilingual BPMN XML translation to English, (ii) execution-oriented validation and correction using a model checker, and (iii) automated process description generation.

\subsection{Foreign language to English model translation}

Translation proceeds in four steps: text extraction, LLM-based translation, fuzzy matching, and XML reconstruction. An XML parser traverses the BPMN file and extracts translatable attributes (e.g., \texttt{name}, \texttt{default}, \texttt{string}) while preserving identifiers. Unique strings are collected to avoid duplicate translations. A structured prompt enforces JSON-formatted input/output during translation. Fuzzy matching resolves minor discrepancies (e.g., encoding or whitespace differences) between extracted strings and translated outputs by selecting the highest-similarity match. The translated content is then reinserted into the XML while preserving structure and formatting.

\subsection{Model correction}
\label{sec: model correction}

Public BPMN diagrams frequently contain compliance errors. We implement a closed-loop correction mechanism in which SpiffWorkflow~\cite{spiffworkflow} provides execution diagnostics that guide LLM-based repairs. Validation enforces execution constraints such as default paths for exclusive gateways and valid condition expressions for non-default branches. For simple diagrams, we regenerate the model; for complex diagrams, we apply localized LLM-proposed repairs (replacement, augmentation, modification, or deletion). The loop iterates until execution compliance is achieved or a repair limit is reached. To control prompt size, only recent interaction history is retained. Visual annotations are temporarily removed during correction and reattached afterward.

\subsection{Process description generation}

Each validated BPMN diagram (with visualization elements removed) is provided to the LLM to generate a well-organized natural language description of the process. The LLM agent explains the process purpose, identifies responsible actors, organizes activities into coherent steps, and translates gateway logic into natural conditional statements. The output consists of concise paragraphs summarizing process flow, participants, and key decision rules.

\section{BPMN Generation}
\label{sec:bpmn generation}
The BPMN generation framework transforms unstructured textual descriptions into executable BPMN~2.0 XML models through a six-stage pipeline, where each stage refines structured outputs from the previous one. The decomposition follows a structured reasoning approach inspired by chain-of-thought methodologies~\cite{10.5555/3600270.3602070} and common practices of human modelers. The pipeline is implemented as a configurable chain-based generator using LangChain~\cite{LangChain}.

\subsection{Process Element Extraction}
The first stage extracts core process components from text, including process boundaries, activities, participants, decisions, inputs, outputs, data flows, and external dependencies. Start and end events define process boundaries, while activities correspond to described tasks and procedural steps. Outputs are returned in structured JSON format to ensure consistency and downstream usability. An excerpt of the prompt used in this stage is shown below.
  \begin{lstlisting}
    You are an expert in process modeling and information extraction.
    Your task is to extract structured process elements from the documentation provided by the user.
    You must identify and extract the following components:
    - **Process boundaries** - What event triggers the start of the process, and what indicates its end.
    - **Activities** - Tasks, actions, or steps involved in the process.
    - **Participants** - Individuals, roles, or entities involved, along with their responsibilities.
    - **Decisions** - Key decision points or gateways affecting the process flow. These can be binary (yes/no) or multi-step.
    - **Inputs** - Required data or materials.
    - **Outputs** - Results or outcomes generated by each activity.
    - **Data flow** - How data moves between steps and is transformed or used.
    - **Dependencies** - Any links to external systems, or processes.
    ...
  \end{lstlisting}

\subsection{Decision Point Analysis}
The second stage formalizes decision logic. For each identified decision point, the system determines required inputs, possible outcomes, and explicit conditions governing each branch. This ensures correct configuration of exclusive gateways, including default paths and condition expressions compatible with execution engine requirements. Outputs maintain a structured mapping between inputs, conditions, and outcomes to support direct translation into BPMN gateway constructs. An excerpt of the prompt used in this stage is shown below.
  \begin{lstlisting}
    You are an expert in process modeling and information extraction.
    Your task is to analyze decision points and determine the input of the decision, the outcomes of the decision and what conditions need to be satisfied for each outcome.
    You must identify and extract the following components:
    - **Decision points** - Required data, key decision points or gateways affecting the process flow.
    - **Inputs** - Required data, information, or prerequisites needed to make each decision. This includes:
      - Data that needs to be checked or evaluated
      - Information that must be available before the decision can be made
      - Status or state that needs to be assessed
      - Results from previous activities that inform the decision
    - **Outputs** - Required data, possible outcomes or results from each decision point.
    - **Conditions** - Required data, specific conditions that need to be satisfied for each output.
    ...
 \end{lstlisting}
\subsection{Data Object Identification}
The third stage identifies and classifies data objects referenced in the process. Objects are categorized as primary (core entities), derived (created during processing), or temporary (used transiently). Each object is described with attributes, properties, relationships, and usage patterns, enabling consistent representation through BPMN data object references and associations. An excerpt of the prompt used in this stage is shown below.
 \begin{lstlisting}
    You are an expert in process modeling and data modeling.
    Your task is to extract **data objects** used by processes, based on information provided by the user.
    To complete this task, perform the following steps:
    - **Identify data objects** relevant to the process.
    - **Classify each data object** as one of: primary (main entity), derived (created during processing), or temporary (used transiently).
    - **Assign clear, descriptive names** to each data object, based on its role or content.
    - **Determine attributes and properties** that describe the data object (e.g., structure, format, units).
    - **Specify how the data object is used**-how it is created, updated, accessed, and archived or deleted.
    - **Describe relationships** between data objects (e.g., containment, linkage, hierarchy).
    - **Explain the flow and transformation** of data objects throughout the process.
    ...
  \end{lstlisting}
\subsection{Data Model Construction}
The fourth stage constructs a structured data model from the identified objects. Entities, attributes, relationships, and constraints (e.g., cardinality, inheritance, primary/foreign keys) are formalized to ensure internal consistency and data integrity. The resulting data architecture supports coherent mapping between process activities and data flows in the generated BPMN model. An excerpt of the prompt used in this stage is shown below.
\begin{lstlisting}
    You are an expert in process modeling and data architecture.
    Your task is to define a structured **data model** for the extracted processes, based on technical documentation and the list of identified entities and data objects provided by the user.
    To accomplish this task, perform the following:
    - **Identify main entities** involved in the process, derived from previously identified data objects.
    - **Determine relationships** between entities, including directionality and type (e.g., references, containment).
    - **List each entity's attributes**, including their data types, constraints, and validation rules.
    - **Specify cardinality and multiplicity** for each relationship (e.g., one-to-many, min/max occurrences).
    - **Include inheritance rules**, where entities inherit properties or behavior from others, if applicable.
    - **Define primary keys, foreign keys, and uniqueness constraints** that enforce data integrity and govern model logic.
    ....
  \end{lstlisting}
\subsection{Data Mapping and Activity Association}

The fifth stage maps activities and decision points to their corresponding data inputs and outputs. For each process element, required inputs, generated outputs, and associated data object properties are explicitly defined. This structured mapping enables generation of BPMN models with complete \texttt{dataInputAssociation} and \texttt{dataOutputAssociation} elements, ensuring execution-level consistency. An excerpt of the prompt used in this stage is shown below.
  \begin{lstlisting}
    You are an expert in process modeling and structured information extraction.
    Your task is to identify **data inputs and outputs** for each activity involved in processes, based on technical documentation, found activities, data objects and models provided by
   the user.

    To complete this task, follow these steps:
    - **Identify each activity** described in the process.
    - For each activity, **determine the required input data**-e.g., user-provided values, database queries, or prior outputs.
    - **Determine the output data** generated or modified by the activity-e.g., calculated values, updated records, or messages sent externally.
    - **Specify the data objects** used as inputs or outputs, including attributes such as format, data type, and validation rules.
    - **Include all elements** of process diagrams (activities such as tasks, events, messages, message events, decision points, such as gateways, etc.)
    - **Make sure** to use only activities and decisions included in the process descriptions.
    - **Make sure** to use only inputs and outputs included in the data objects and model.
    ...
  \end{lstlisting}

\subsection{BPMN XML Generation}
The final stage synthesizes all structured artifacts into executable BPMN~2.0 XML. Generation enforces specification compliance, excludes BPMNDI visualization elements to reduce model size, and ensures inclusion of all identified activities and gateways. Execution engine constraints include: (i) default paths for exclusive gateways, (ii) condition expressions for non-default branches, (iii) correct ordering of data object references, and (iv) complete sequence-flow connectivity. Generated models are validated using SpiffWorkflow and XML parsing utilities to ensure namespace, syntax, and structural correctness. When validation fails, models enter the correction loop described in Section~\ref{sec: model correction}. BPMNDI annotations are added after validation for visualization in tools such as Camunda. An excerpt of the prompt used in this stage is shown below.
  \begin{lstlisting}
    You are a BPMN modeling expert with deep knowledge of syntax and execution semantics for tools like Camunda.
    Your task is to generate a syntactically correct BPMN 2.0 XML diagram using only structured input provided by the user.
    The input will consist of structured dictionaries that define:
    - Data objects and their properties
    - Data inputs and outputs per activity
    - Participants (roles or entities)
    - Decision points (gateways)
    - Relationships between entities (if needed to define subprocesses or swimlanes)
    The requirements for the diagram are:
    - The diagram respects the BPMN specs.
    - Do NOT add BPMNDI annotations for visualization purposes to limit the size of the model.
  ...
  \end{lstlisting}

\section{BPMN model comparison}
\label{sec:comparison}
BPMN diagram comparison can be seen from the lens of a graph isomorphism problem which is a challenging combinatorial problem \cite{10.1145/3372123}, at most solvable in quasipolynomial time \cite{10.1145/2897518.2897542}. Metrics based on features or graph kernels (e.g., the Weisfeiler–Lehman subtree kernel) \cite{JMLR:v12:shervashidze11a} compare ordered label multiset features, but struggle when node/edge labels vary or orders differ. Embedding-based methods and semantic similarity help overcome this by mapping structural and label signals onto continuous space, where similarity is judged more flexibly even when feature orders or label vocabularies differ. In this section we describe a scoring mechanism that evaluates BPMN similarity across multiple dimensions. The assessment combines structural analysis and semantic understanding to produce similarity metrics. The mechanism integrates graph-theoretic measures with natural language processing techniques.

We organize BPMN elements into five primary categories: tasks, gateways, events, data elements, and flow elements. This classification focuses on functional semantics rather than syntactic differences, allowing, for example, user tasks and service tasks to be compared within the same category. To operationalize these metrics, we utilize the NetworkX library for structural graph analysis and the {\tt all-MiniLM-L6-v2} sentence-transformer model for generating semantic embeddings. This combination allows us to compute similarity scores that reflect both the topological integrity and the semantic intent of the generated models.

\subsection{Structural Similarity}
The structural similarity dimension evaluates topological connectivity patterns between BPMN models through graph analysis. The assessment considers multiple structural characteristics including node count, edge count, graph density, average degree, and degree sequence patterns. The topological similarity computation uses basic graph statistics, where similarity for each metric is calculated using the ratio formula:
\begin{equation}
S_{metric} = \frac{\min(M_1, M_2)}{\max(M_1, M_2)},
\end{equation}
where $M_1$ and $M_2$ represent the metric values for the two graphs being compared. This formulation generates a symmetric similarity score in the range [0,1]. Degree sequence similarity employs correlation analysis to assess structural pattern correspondence. The degree sequences are extracted from both graphs, padded to equal length, and then compared using Pearson correlation coefficient:
\begin{equation}
S_{degree} = |\rho(D_1, D_2)|,
\end{equation}
where $D_1$ and $D_2$ represent the degree sequences and $\rho$ denotes the correlation coefficient. The structural similarity score is the mean of individual metric scores.

\subsection{Type Distribution Similarity}
The type distribution dimension assesses the compositional similarity between BPMN models by analyzing the distribution of element types. This analysis recognizes that functionally similar processes should exhibit comparable distributions of different element types, even if the specific connectivity patterns differ. The assessment employs Jensen-Shannon divergence to quantify distributional differences between type frequency patterns. The type counts are first normalized to probability distributions:
\begin{equation}
P_1(t) = \frac{C_1(t)}{\sum_{t'} C_1(t')}, \quad P_2(t) = \frac{C_2(t)}{\sum_{t'} C_2(t')}
\end{equation}
where $C_i(t)$ represents the count of type $t$ in graph $i$. The Jensen-Shannon divergence is computed using the symmetric formulation:
\begin{equation}
JS(P_1, P_2) = \frac{1}{2} D_{KL}(P_1 || M) + \frac{1}{2} D_{KL}(P_2 || M)
\end{equation}
where $M = \frac{1}{2}(P_1 + P_2)$ is the mixture distribution and $D_{KL}$ denotes Kullback-Leibler divergence. Type-distribution similarity is computed as:
\begin{equation}
S_{type\_dist} = \max(0, 1 - JS(P_1, P_2))
\end{equation}
ensuring the similarity score remains within the valid range [0,1].

\subsection{Semantic Similarity}
The semantic dimension leverages natural language processing techniques to evaluate the similarity of text. This assessment recognizes that process models with different terminologies may represent functionally equivalent activities, requiring semantic understanding beyond exact string matching. The semantic analysis employs pre-trained sentence transformer models \cite{reimers-2019-sentence-bert} to generate embedding vector representations of textual content. The similarity computation utilizes cosine similarity between embedding vectors to capture semantic relationships. We use optimal assignment matching to ensure maximum similarity alignment between text elements. This approach addresses the challenge of determining which elements in one model correspond to which elements in the comparison model, maximizing the overall semantic correspondence.
The semantic similarity analysis creates label representations that incorporate neighborhood information. For each node, the context string is constructed as:
\begin{equation}
\text{context}(n) = \text{label}(n) \; || \; \text{"neighbors: "} \; || \; \bigcup_{m \in N(n)} \text{label}(m)
\end{equation}
where $N(n)$ represents the set of neighboring nodes and $||$ denotes string concatenation. The neighbor labels are sorted alphabetically to ensure consistent representation regardless of processing order. The semantic similarity analysis produces three distinct similarity measures: element \textit{name/description} similarity, \textit{type} similarity and \textit{combined} similarity using merged name-type representations. The \textit{combined} similarity employs concatenated \textit{name-type} strings to capture both semantic content and structural role information simultaneously.
The final \textit{overall similarity} score for one reconstruction experiment is the \textit{average across all five similarity-metric dimensions}.

\section{Experimental results}
\label{sec:experimental results}
The experimental framework employs a two-phase evaluation strategy consisting of (i) model reconstruction %execution
and (ii) quantitative similarity assessment against curated ground-truth references. The reconstruction pipeline proceeds through several stages: (a) collection of open-source BPMN models, (b) translation of models into English, (c) SpiffWorkflow-based structural validation combined with LLM-driven iterative correction to establish ground-truth models, (d) LLM-based generation of process descriptions, and (e) LLM-based reconstruction of BPMN models. The resulting reconstructions are then compared with the ground-truth models using the methodology outlined in Section~\ref{sec:comparison}.

A total of 750 models were initially collected and processed through translation, correction, and description generation using {\tt chatGPT-4o}, resulting in 387 validated models. The collected dataset ensures domain diversity, spanning healthcare (clinical pathways), finance (loan approvals), and supply chain logistics. We then conducted three reconstruction experiments employing {\tt chatGPT-4o}, {\tt gemini-2.5-flash}, and {\tt gemini-2.5-pro}. Table~\ref{tab:llm-reconstruction} reports the number of {\tt SpiffWorkflow}-compliant models that each LLM was able to reconstruct from textual descriptions. The {\tt gemini} family achieved the strongest results, with the {\tt pro} variant successfully reconstructing all models.

\begin{table}[t]
\centering
\caption{LLM reconstruction count.}
\label{tab:llm-reconstruction}
\begin{tabular}{l|c|c|c}
\hline
\textbf{Descriptions} & \textbf{chatGPT-4o} & \textbf{gemini-2.5-flash} & \textbf{gemini-2.5-pro}\\
\hline
387 & 353 & 370 & 387\\
\hline
\end{tabular}
\end{table}

We next evaluated the similarity metrics for the models reconstructed by the three LLMs. The quality of reconstruction was influenced both by the choice of LLM and by the amount of information available in the process descriptions. When descriptions lacked sufficient detail, the LLMs tended to interpolate missing elements, which introduced discrepancies in component names and textual labels relative to the ground truth. Logical correctness of gateway expressions was not evaluated and remains future work.

\begin{figure}[htp!]
    \centering
    \includegraphics[width=0.9\linewidth]{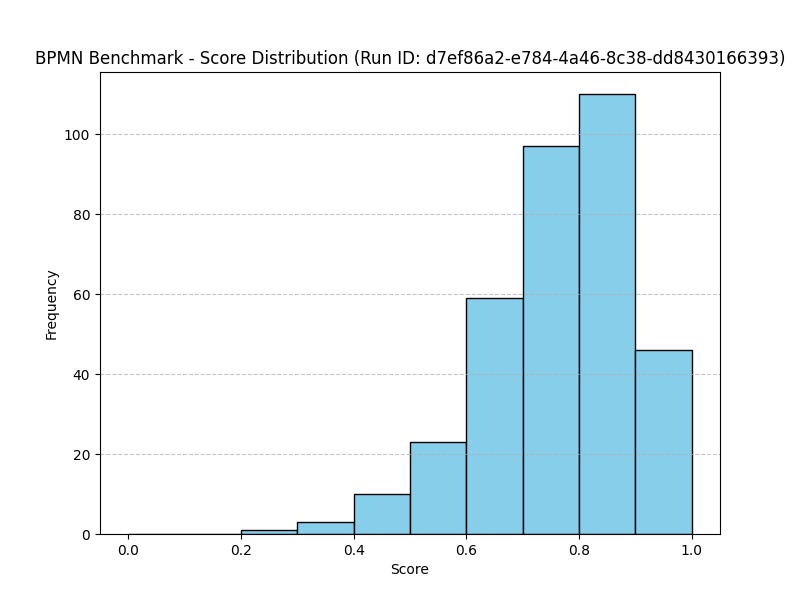}
    \caption{Score distribution histogram: {\tt chatGPT-4o}. }
    \label{fig:score_distribution}
\end{figure}

\begin{figure}[htp!]
    \centering
    \includegraphics[width=0.9\linewidth]{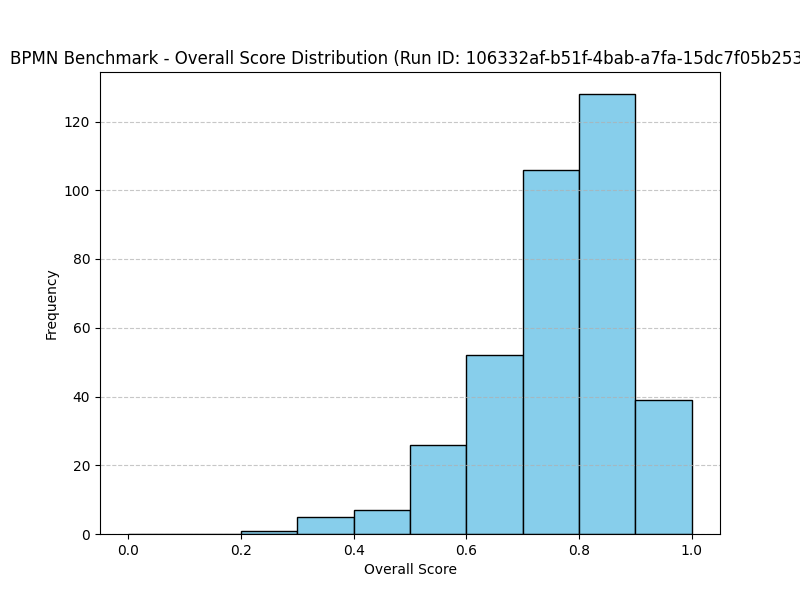}
    \caption{Score distribution histogram: {\tt gemini-2.5-flash}.}
    \label{fig:score_distribution_flash}
\end{figure}

\begin{figure}[htp!]
    \centering
    \includegraphics[width=0.9\linewidth]{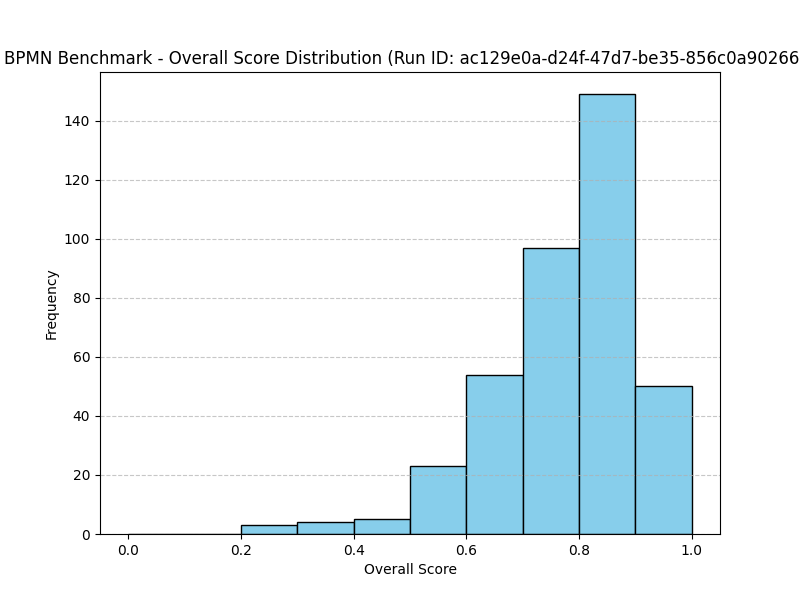}
    \caption{Score distribution histogram: {\tt gemini-2.5-pro}.}
    \label{fig:score_distribution_pro}
\end{figure}

Figures~\ref{fig:score_distribution}, \ref{fig:score_distribution_flash}, and \ref{fig:score_distribution_pro} present the histograms of overall similarity scores for the three LLMs, while Tables~\ref{tab:evaluation-averages-4o}, \ref{tab:evaluation-averages-flash}, and \ref{tab:evaluation-averages-pro} provide the corresponding detailed results. {\tt chatGPT-4o} and {\tt gemini-2.5-flash} achieve comparable performance, whereas {\tt gemini-2.5-pro} shows a small improvement, additionally achieving approximately 50 near-perfect reconstructions. The reported scores are normalized by the number of successfully reconstructed models; when normalized by the number of process descriptions, the {\tt gemini} family outperforms {\tt chatGPT-4o} by a larger margin. While {\tt gemini-2.5-pro} model improves reconstruction quality, often taking only one iteration to fix errors, it does incurs higher latency (up to five minutes per model).  For instance, the overall average score would decrease to 0.6981 for {\tt chatGPT-4o} and 0.7350 for {\tt gemini-2.5-flash}. Despite the absence of manual curation, the results demonstrate strong baseline performance for fully automated BPMN reconstruction. 

\begin{table}[t]
\centering
\caption{Average evaluation scores across similarity dimensions: {\tt chatGPT-4o}.}
\label{tab:evaluation-averages-4o}
\begin{tabular}{lcc}
\hline
\textbf{Metric} & \textbf{Average Score} \\
\hline
Structural Similarity        & 0.8050  \\
Type Distribution Similarity & 0.9112  \\
Name/Description Semantic Similarity           & 0.6166  \\
Type Semantic Similarity            & 0.7944  \\
Name-Type Semantic Similarity        & 0.6997  \\
Overall Similarity              & 0.7654 \\
\hline
\end{tabular}
\end{table}

\begin{table}[t]
\centering
\caption{Average evaluation scores across similarity dimensions: {\tt gemini-2.5-flash}.}
\label{tab:evaluation-averages-flash}
\begin{tabular}{l c}
\hline
\textbf{Metric} & \textbf{Average Score} \\
\hline
Structural Similarity       & 0.8098 \\
Type Distribution Similarity & 0.9033 \\
Name/Description Semantic Similarity          & 0.6256 \\
Type Semantic Similarity           & 0.8047 \\
Name-Type Semantic Similarity       & 0.7006 \\
Overall Similarity        & 0.7688 \\
\hline
\end{tabular}
\end{table}

\begin{table}[t]
\centering
\caption{Average evaluation scores across similarity dimensions: {\tt gemini-2.5-pro}.}
\label{tab:evaluation-averages-pro}
\begin{tabular}{l c}
\hline
\textbf{Metric} & \textbf{Average Score} \\
\hline
Structural Similarity       & 0.8173 \\
Type Distribution Similarity & 0.9295 \\
Name/Description Semantic Similarity          & 0.6255 \\
Type Semantic Similarity           & 0.8104 \\
Name-Type Semantic Similarity       & 0.7021 \\
Overall Similarity        & 0.7770 \\
\hline
\end{tabular}
\end{table}

We analyzed examples of poor reconstruction results (similarity $<0.5$) to understand failure modes. Failures primarily stemmed from three sources: 1) Ambiguous Branching Logic: When descriptions lacked explicit conditions for exclusive gateways, the LLM often hallucinated arbitrary logic to satisfy the model's execution validity requirements. 2) Implicit Dependencies: The LLM sometimes failed to capture error-handling paths that were implied by context but not explicitly stated in the text. 3) Abstraction Mismatches: In some cases, the LLM aggregated multiple low-level steps into a single high-level activity, creating structural divergence from the ground truth despite maintaining semantic accuracy. An example of process description originating from a syntactically correct  BPMN model that is logically incomplete is shown bellow.
  \begin{lstlisting}
  ...
 The diagram does not provide further details on the tasks or activities that follow the parallel paths, nor does it specify any conditions or criteria for the completion of these tasks. The focus is primarily on the initiation of the process and the immediate branching into parallel paths, leaving the specifics of subsequent activities open to interpretation or further development.
 ...
  \end{lstlisting}

\section{Related Work}
\label{sec:related work}

Research on automatic BPMN generation from text spans more than a decade, evolving from rule-based approaches to LLM-driven pipelines. Friedrich et al.~\cite{friedrich2011automated} pioneered syntactic parsing and template-based transformations, but faced semantic ambiguity and limited executability. Bellan et al.~\cite{bellan2021process} benchmarked extraction methods and highlighted shortcomings in datasets and evaluation practices. Qayyum et al.~\cite{qayyum2023dialogue} proposed dialogue-driven derivation of task relationships for rapid prototyping, but did not produce executable BPMN~2.0 XML; Zirnstein~\cite{zirnstein2024extraction} studied extraction from unstructured text. In contrast, we construct ground truth at scale from heterogeneous repositories and enforce execution-oriented compliance via SpiffWorkflow.

Recent work leverages LLMs for interactive assistance and information extraction. K\"opke and Safan~\cite{koepke2024bpmnchatbot}, Kourani et al.~\cite{kourani2024promoai}, and H\"orner~\cite{horner2025outlines} emphasize usability and interactive modeling rather than large-scale benchmarking. Neuberger et al.~\cite{neuberger2024universal} focus on universal prompting for mention detection, entity resolution, and relation extraction, treating BPMN generation as a proof of concept. Adjacent directions include simulation-model synthesis from logistics/manufacturing text~\cite{monti2024introduces}, process-description generation and verification~\cite{nivon2025givup}, structured plan representations~\cite{garg2025sopstruct}, multimodal extraction~\cite{voelter2024multimodal}, and parallelism detection~\cite{le2025leveraging}. Our focus is end-to-end automation with enforcement of specification compliance and large-scale evaluation.

Several systems rely on curated datasets or constrained inputs. Held~\cite{Held2023} uses gold standards and regulatory-text preprocessing and evaluates component-level extraction accuracy. BPMN Sketch Miner~\cite{10.1145/3365438.3410990} uses constrained natural language during live modeling for incremental diagram synthesis. Zubenko~\cite{Zubenko2024} applies prompt-engineering with iterative correction for syntactic issues, often regenerating models when failures occur. We instead use SpiffWorkflow as a model checker to validate execution-oriented constraints and scale to larger processes by applying localized, programmatic repairs rather than wholesale regeneration.

A key distinction between our work and prior approaches lies in the evaluation. Early systems such as Friedrich et al.~\cite{friedrich2011automated} emphasized structural validity, while recent LLM frameworks~\cite{kourani2024promoai,horner2025outlines,koepke2024bpmnchatbot} often rely on qualitative studies. Dataset-centric work~\cite{bellan2021process,neuberger2024universal} typically reports task-level extraction accuracy. Klievtsova et al.~\cite{klievtsova2024conversationalprocessmodelinggenerative} propose text-to-task overlap-style metrics. Because we operate directly on BPMN files, we evaluate both topology and semantics (via embeddings) at the model level, with explicit attention to compliance and repair.

\section{Conclusions and future developments}
\label{sec:conclusions}
This paper presented an LLM-driven framework for reconstructing BPMN~2.0 models from textual process descriptions. The methodology integrates automated ground-truth generation, description synthesis, and staged reconstruction. A multi-dimensional similarity framework was introduced, combining structural, semantic, and contextual measures to provide a principled basis for evaluation. Experiments on 750 publicly sourced BPMN diagrams yielded approximately 400 validated ground-truth models, with average similarity scores up to 0.77 and 50 near-perfect cases. The results demonstrate scalable BPMN reconstruction without manual curation and robust structural interpolation under incomplete specifications. While our approach achieves structural executability — producing valid gateways, flows, and condition expressions — fully operationalizing these models requires task-level implementation details. Future work will focus on enriching these models with executable payloads, such as generating Python snippets for script tasks and configuring API connectors for service tasks.  This will bridge the final gap between the structurally executable XML we currently generate and fully autonomous, production-ready process workflows.

\bibliographystyle{plain}
\bibliography{bibtex/ref}  

\end{document}